\documentclass[aps,prl,showpacs,nofootinbib,twocolumn,
nobalancelastpage,superscriptaddress]{revtex4-1}
\usepackage[dvips]{epsfig}
\usepackage{graphicx}
\usepackage{amsmath}    
\usepackage{color}

\newcommand{\postscript}[2]{\setlength{\epsfxsize}{#2\hsize}
   \centerline{\epsfbox{#1}}}

\newcommand{\epm}{\ensuremath{e^{\pm}\;}}
\begin{document}

%%%%%%%%%%%%%%%%%%%%%%%%%%%%%%%%%%%%%%%%%%%%%%%%%%%%%%%%%%%%%%%

\title{Right-Handed Neutrinos as the Dark Radiation: Status and Forecasts for the LHC}

\author{Luis A.  Anchordoqui} 
\affiliation{Department of Physics,
  University of Wisconsin-Milwaukee, Milwaukee, WI 53201}
\author{Haim  Goldberg}
\affiliation{Department of Physics,
Northeastern University, Boston, MA 02115}

\author{Gary Steigman} \affiliation{Center for Cosmology and
  Astro-Particle Physics, Ohio State University, Columbus, OH 43210} \affiliation{Department of Physics, Ohio State University,
  Columbus, OH 43210} \affiliation{Department of Astronomy, 
  Ohio State University, Columbus, OH 43210}

\date{November 2012}

\begin{abstract}
  \noindent Precision data from cosmology (probing the CMB decoupling
  epoch) and light-element abundances (probing the BBN epoch) have
  hinted at the presence of extra relativistic degrees of freedom, the
  so-called ``dark radiation.''  We present a model independent study
  to account for the dark radiation by means of the right-handed
  partners of the three, left-handed, standard model neutrinos. We
  show that milli-weak interactions of these Dirac states (through
  their coupling to a TeV-scale $Z'$ gauge boson) may allow the
  $\nu_R$'s to decouple much earlier, at a higher temperature, than
  their left-handed counterparts.  If the $\nu_R$'s decouple during
  the quark-hadron crossover transition, they are considerably cooler
  than the $\nu_L$'s and contribute less than 3 extra ``equivalent
  neutrinos'' to the early Universe energy density. For decoupling in
  this transition region, the 3\,$\nu_R$ generate $\Delta N_{\nu} =
  3(T_{\nu_R}/T_{\nu_L})^{4} < 3$, extra relativistic degrees of
  freedom at BBN and at the CMB epochs. Consistency with present
  constraints on dark radiation permits us to identify the allowed
  region in the parameter space of $Z'$ masses and
  couplings. Remarkably, the allowed region is within the range of
  discovery of LHC14.
\end{abstract}

\maketitle

%%%%%%%%%%%%%%%%%%%%%%%%%%%%%%%%%%%%%%%%%%%%%%%%%%%%%%%%%%%%%%%

Big-bang nucleosynthesis (BBN) is remarkably succesfull in predicting
the relative abundance of light elements as a function of two
fundamental parameters: the baryon density of the universe,
$\Omega_{\rm B} h^2$, and the number of ``equivalent'' light neutrino
species, $N_{\rm eff}$~\cite{Steigman:1977kc}.\footnote{The discussion
  here is in the context of the usual concordance cosmology of a flat
  universe dominated by a cosmological constant, with $\Omega_\Lambda
  \sim 0.7$ and a cold dark matter plus baryon component $\Omega_m
  \sim 0.3$; the Hubble parameter as a function of redshift is given
  by $H^2(z) = H_0^2 [\Omega_m (1+z)^3 + \Omega_\Lambda]$, normalized
  to its value today, $H_0 \sim 100~h\ {\rm km} \ {\rm s}^{-1} \ {\rm
    Mpc}^{-1}$, with $h \simeq 0.71$~\cite{Beringer:1900zz}.} In fact,
until recently BBN provided the only constraint on these parameters.
Discovery of primordial anisotropies in the cosmic microwave
background (CMB) has granted a superior test-bed for precision
constraints on fundamental parameters in cosmology. This powerful
test-bed can be used to assess whether new physics model predictions
are simultaneously consistent with BBN and CMB observations. Of
interest here is the capacity to probe right-handed neutrino
milli-weak
interactions~\cite{Ellis:1985fp,GonzalezGarcia:1989py,Lopez:1989dh,Barger:2003zh,Anchordoqui:2011nh},
which are predicted in various extensions of the standard model of
particle physics.

In this Letter we construct a model independent template for placing
upper and lower bounds on the mass of an extra $Z'$ gauge boson, which
allows for milli-weak interactions of the right-handed partner of the
Dirac neutrino.  A critical input for such an analysis is the relation
between the relativistic degrees of freedom (r.d.o.f.)  and the
temperature of the primordial plasma. This relation is complicated
because the temperature which is of interest for right-handed neutrino
decoupling from the heat bath may lay in the vicinity of the
quark-hadron cross-over transition. In a previous
publication~\cite{Anchordoqui:2011nh}, use was made of a detailed
lattice study to connect the temperature to an effective number of
degrees of freedom. Very recently, one of us has provided an analysis
in which the decoupling of the extra relativistic degrees of freedom
may occur well beyond the cross-over
temperature~\cite{Steigman:2012ve,Steigman:2012nb}. In that case a general connection
between the effective number of degrees of freedom and the
right-handed neutrino decoupling temperature is needed.  To this end,
we employ the results of Ref.~\cite{Laine:2006cp} to find the
post-\epm annihilation ratio of the temperatures of the right-handed
and left-handed neutrinos, $T_{\nu_R}/T_{\nu_L}$, which is then used to
predict the enhancement to the effective number of degrees of freedom
in the early Universe, $\Delta N_{\nu} = 3(T_{\nu_R}/T_{\nu_L})^{4} <
3$.

The formulation presented here allows for an immediate test for the
potential of any model to account for any extra neutrino degrees of
freedom. For illustration, we analyze several candidate models. Before
proceeding we provide a brief and concise overview of the current
observational constraints on the number of light neutrino species.

Over the past few years, the Wilkinson Microwave Anisotropy Probe
(WMAP)~\cite{Komatsu:2010fb}, the Atacama Cosmology Telescope
(ACT)~\cite{Dunkley:2010ge}, and the South Pole Telescope
(SPT)~\cite{Keisler:2011aw} have each provided evidence for a ``dark''
relativistic background (a.k.a. dark radiation). Parameterized in
terms of the number of relativistic degrees of freedom the data seem
to favor the existence of roughly one extra effective neutrino
species.  Specifically, the parameter constraint from the combination
of WMAP 7-year data, the latest distance measurements from the baryon
acoustic oscillations (BAO) in the distribution of
galaxies~\cite{Percival:2009xn}, and precise measurements of
$H_0$~\cite{Riess:2009pu} lead to $N_{\rm eff} = 4.34^{+0.86}_{-0.88}
\, (68\% {\rm CL})$~\cite{Komatsu:2010fb}.  Similarly, a combination
of BAO and $H_0$ with data from the ACT yields $N_{\rm eff} = 4.56 \pm
0.75 \, (68\% {\rm CL})$~\cite{Dunkley:2010ge}, whereas data collected
with the SPT combined with BAO and $H_0$ arrive at $N_{\rm eff} = 3.86
\pm 0.42\, (1\sigma)$~\cite{Keisler:2011aw}.

Turning now to the BBN determinations, we note that the
observationally-inferred primordial fractions of baryonic mass in
$^{4}$He had been favoring $N_{\rm eff} \alt 3$~\cite{Simha:2008zj}.
Unexpectedly, two recent independent studies determined a larger
$^4$He fraction and the updated effective number of light neutrino
species is reported as $N_{\rm eff} = 3.80^{+0.80}_{-0.70}$
($2\sigma$)~\cite{Izotov:2010ca,Aver:2010wq}.

\begin{figure}[tbp]
\postscript{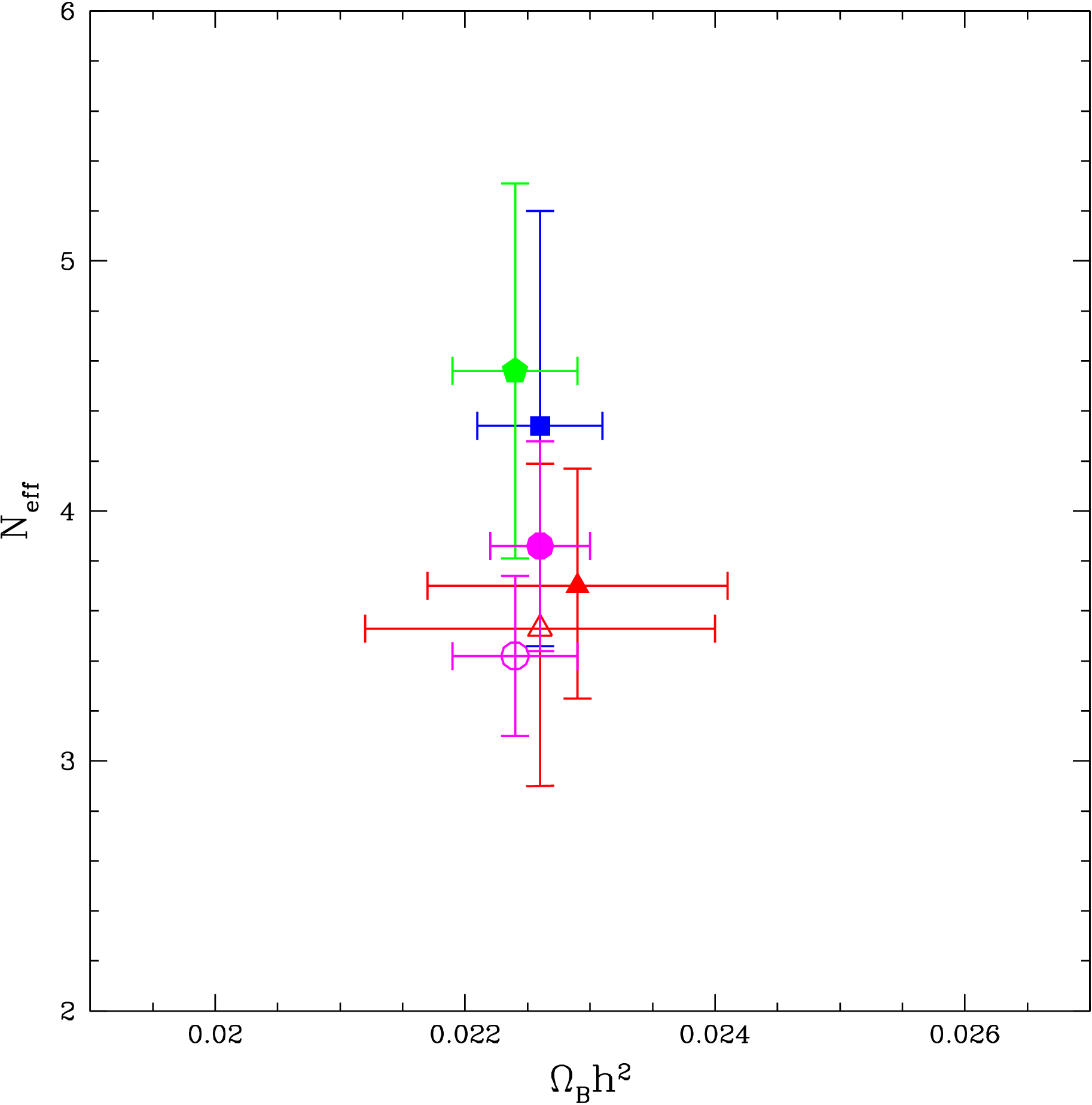}{0.99}
\caption{Comparing the BBN predictions of $N_{\rm eff}$ and
  $\Omega_{\rm B} h^2$ with those from various CMB determinations: BBN
  D + $^4$He (red filled triangle)~\cite{Steigman:2012ve}, BBN D +
  WMAP7~\cite{Komatsu:2010fb} (red open triangle),
  WMAP7~\cite{Komatsu:2010fb} (blue filled square),
  ACT~\cite{Dunkley:2010ge} (green filled pentagon),
  SPT~\cite{Keisler:2011aw} (purple filled circle), SPT +
  Clusters~\cite{Hou:2011ec} (purple open circle). Taken from
  Ref.~\cite{Steigman:2012ve}.}
\label{fig:ggi_1}
\end{figure}

Several recent papers have presented results from combined analysis of
the BBN and/or CMB data, but based on different priors (see
e.g.~\cite{Hou:2011ec,Archidiacono:2011gq,Hamann:2011hu,Nollett:2011aa,Moresco:2012by}).
Throughout we adopt $N_{\rm eff} = 3.71 ^{+0.47}_{-0.45}$ ($1\sigma$),
which represents a conservative choice based on an extensive BBN
analysis~\cite{Steigman:2012ve}. As shown in
Ref.~\cite{Steigman:2012ve} and display here in Fig.~\ref{fig:ggi_1},
this analysis is in agreement with various CMB observations.  In the
standard electroweak theory, $N_{\rm eff} \simeq 3.046$ (the
difference from 3 being mainly due to partial heating of $\nu_L$ by
$e^+e^-$ annihilation~\cite{Mangano:2005cc}), yielding
\begin{equation}
\Delta N_\nu =
0.66^{+0.47}_{-0.45} \ (1\sigma) \ .
\label{deltanu}
\end{equation}

%%%%%%%%%%%%%%%%%%%%%%%
%%%%%%%%%%%%%%%%%%%%%%%%%

Several explanations have been proposed to explain such a possible
$\Delta N_\nu$ excess. These include: {\it (i)} models in which the
extra relativistic degrees of freedom are related to possible dark
matter
candidates~\cite{Ichikawa:2007jv,Hasenkamp:2011em,Menestrina:2011mz,Feng:2011in,Hooper:2011aj,Bjaelde:2012wi};
{\it (ii)} models based on active-sterile mixing of neutrinos in a
heat bath~\cite{Hamann:2010bk,Krauss:2010xg}; {\it (iii)} models based
on milli-weak interactions of right-handed partners of three Dirac
neutrinos~\cite {Anchordoqui:2011nh}. In this work we confine our
discussion to case ${\it (iii)}$.

\begin{figure}[tbp]
\postscript{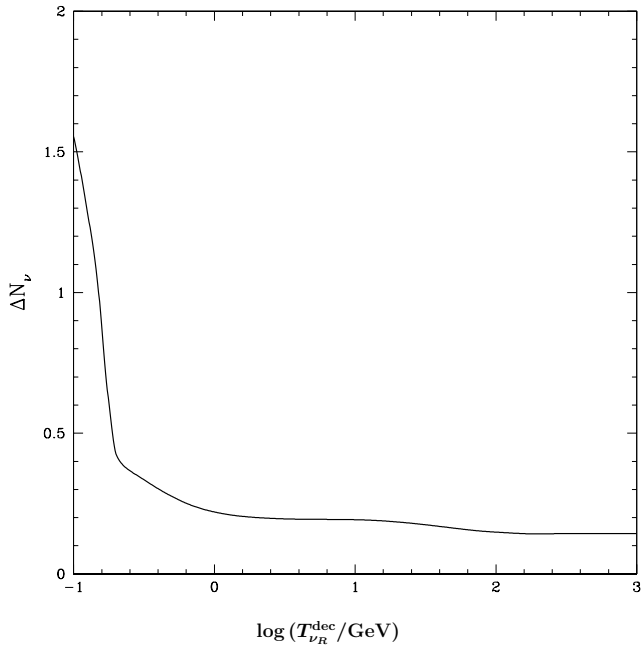}{0.99}
\caption{Relation between $\Delta N_\nu \ {\it vs.} \ T_{\nu_R}^{\rm
  dec}$.}
\label{fig:ggi_2}
\end{figure}

We begin by first establishing, in a model independent manner, the
range of decoupling temperatures implied by  the BBN or  CMB
observations. The effective number of
neutrino species contributing to r.d.o.f. can be written as $
N_{\rm eff} = 3 [1 + (T_{\nu_R}/T_{\nu_L}) ^4] \,;$ therefore,
taking into account the isentropic heating of the rest of the plasma
between $T_{\nu_R}^{\rm dec}$ and $T_{\nu_L}^{\rm dec}$ decoupling
temperatures we obtain
\begin{equation}
\Delta N_\nu = 3 \left(\frac{g(T_{\nu_L}^{\rm dec})}{g(T_{\nu_R}^{\rm
      dec})} \right)^{4/3} \, ,
\label{7}
\end{equation}
where $g(T)$ is the effective number of interacting (thermally
coupled) r.d.o.f. at temperature $T$; for example, $g(T_{\nu_L}^{\rm
  dec}) = 43/4$~\cite{Kolb:1990vq}.\footnote{ If relativistic
  particles are present that have decoupled from the photons, it is
  necessary to distinguish between two kinds of $g$: $g_\rho$ which is
  associated with the total energy density, and $g_s$ which is
  associated with the total entropy density.  For our calculations we use $g = g_\rho = g_s$.}
For the particle content of the standard model, there is a maximum of
$g(T_{\nu_R}^{\rm dec}) = 427/4$ (with $T_{\nu_R}^{\rm dec} > m_{\rm
  top}$), which corresponds to a minimum value of $\Delta N_\nu =
0.14$. For the subsequent study, we adopt the determination of $g(T)$
given in~\cite{Steigman:2012nb} based on the results
of~\cite{Laine:2006cp}. Then using Eq.~(\ref{7}) we obtain a relation
between $\Delta N_\nu \ {\it vs.} \ T_{\nu_R}^{\rm dec}$, which is
shown in Fig.~\ref{fig:ggi_2}.  From this curve we determine the range
of decoupling temperature: $T_{\nu_R}^{\rm dec} =
0.174^{+1.326}_{-0.030}~{\rm GeV}$.

\begin{figure*}[tbp]
\begin{minipage}[t]{0.49\textwidth}
\postscript{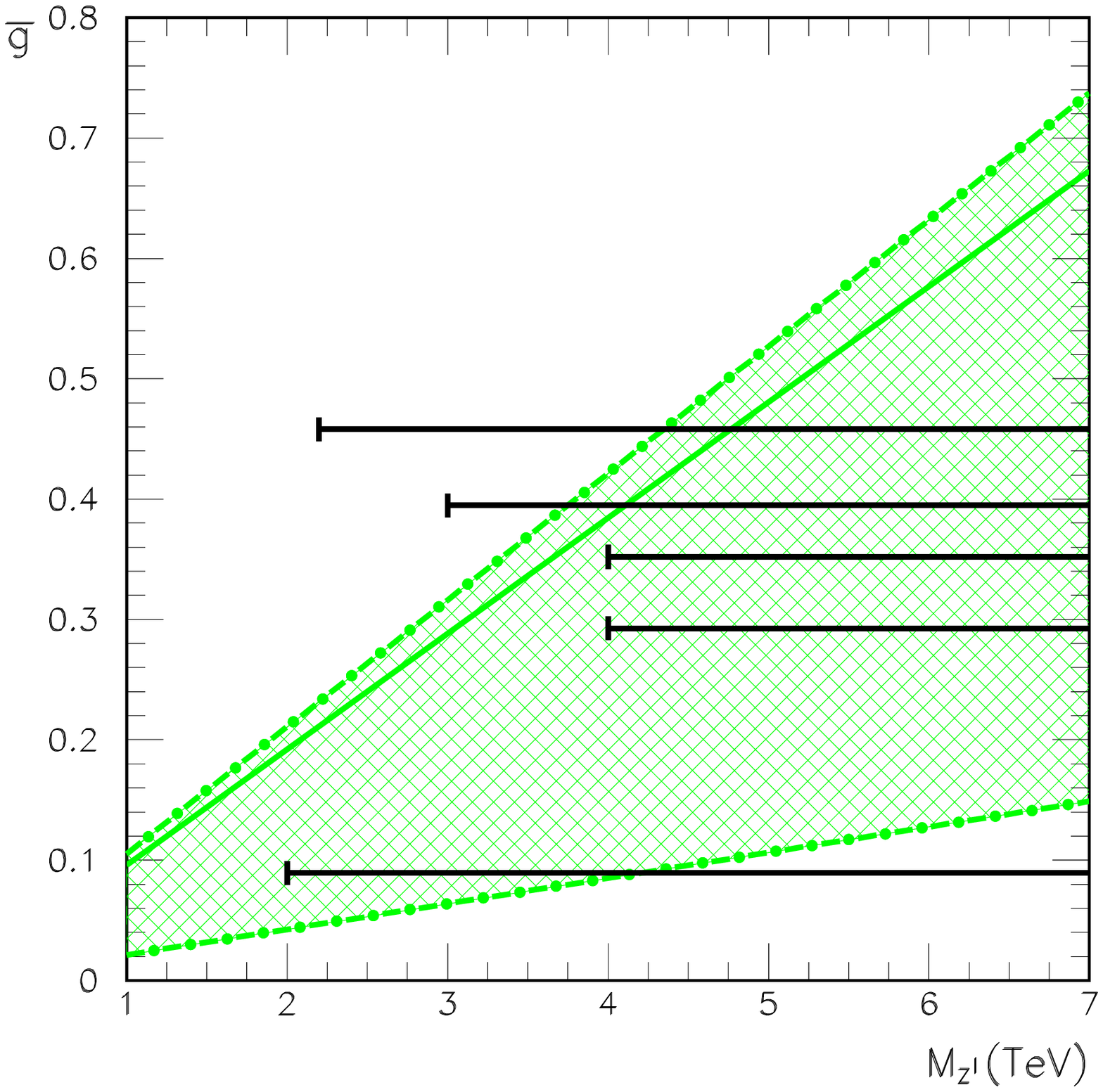}{0.99}
\end{minipage}
\hfill
\begin{minipage}[t]{0.49\textwidth}
\postscript{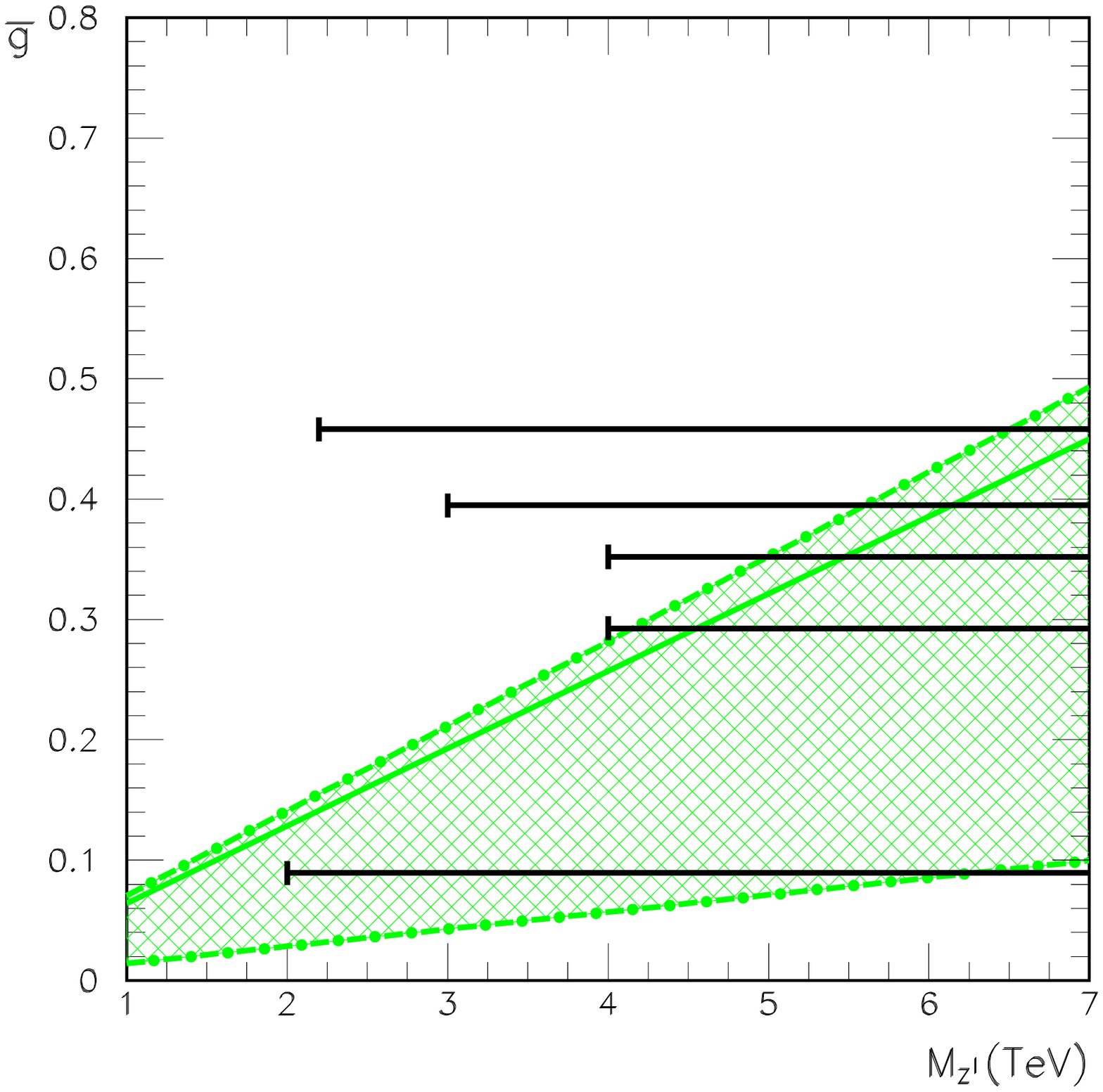}{0.99}
\end{minipage}
\caption{The green cross-hatched areas show the region allowed from
  decoupling requirements to accommodate BBN and CMB eras. Each of the
  horizontal lines refers to a particular model: from the top the
  $E_6$ $Z'_\chi$, a
  D-brane model in which $Z'$ is mostly $B-L$, a
  D-brane model in which $Z'$ couples mostly to the third component of
  a right-handed isospin, a D-brane model with TeV-scale strings, the
  $E_6$ $Z'_\psi$. Termination of the lines on the left reflects the LHC
  experimental limits on the mass of the gauge boson. The left and
  right figures show the condition on decoupling for loss of chemical
  and thermal equilibrium, respectively.}
\label{fig:ggi_3}
\end{figure*}

The physics of interest then will be taking place at energies in the
region of the quark-hadron crossover transition, so that we will
restrict ourselves to the following fermionic fields, and their
contribution to r.d.o.f.: $ \left[ 3 u_R \right] + \left[ 3 d_R
\right] + \left[ 3s_R\right ] + \left[ 3 \nu_L + e_L + \mu_L \right] +
\left[ e_R + \mu_R \right]+ \left[ 3 u_L + 3d_L + 3s_L\right]+ \left[
  3 \nu_R \right]$. This amounts to 28 Weyl fields, translating to 56
fermionic r.d.o.f.\footnote{In principle, the contributions to the
  $\nu_R$ interaction rate $\Gamma(T)$ from the $c$ quark and $\tau$
  lepton should be included for the lowest value of $\Delta N_\nu$,
  which in this paper corresponds to a decoupling temperature of
  1.5~GeV. However, one can easily verify that this inclusion will not
  be visible in Fig.~\ref{fig:ggi_3}.}

%%%%%%%%%%%%%%%%%%%%%%%
%%%%%%%%%%%%%%%%%%%%%%%%%

%%%%%%%%%%%%%%%%%%%%%%%
%%%%%%%%%%%%%%%%%%%%%%%%%

The right-handed neutrino decouples  from the plasma when its mean
free path becomes greater than the Hubble radius at that time.  To
determine $T_{\nu_R}^{\rm dec}$, we first calculate the $\nu_R$ interaction rate 
\begin{equation}
\Gamma (T) = {\cal K} \ \frac{1}{8} \ \left(\frac{\overline  g}{M_{Z'}}
\right)^4 \ T^5 \ \sum_{i=1}^6 {\cal N}_i \,,
\label{Gamma}
\end{equation}
where ${\cal N}_i$ is the number of chiral states,
\begin{equation}
\overline g \equiv 
\left(\frac{\sum_{i =1}^6  {\cal N}_i g_i^2 g_6^2}{\sum_{i =i}^6  {\cal
    N}_i} \right)^{1/4}\, ,
\label{gbarra}
\end{equation}
$g_i$ are the chiral couplings of the $Z'$ gauge boson, and the
cosntant ${\cal K}
= 0.5 \ (2.5)$ for annihilation (annihilation + scattering)~\cite {Anchordoqui:2011nh}.
Armed with (\ref{Gamma}) we determine $T_{\nu_R}^{\rm dec}$  via the prescription
\begin{equation}
 \Gamma(T_{\nu_R}^{\rm dec}) = H(T_{\nu_R}^{\rm dec}) \, ,
\label{haim1}
\end{equation}
where
\begin{equation}
H(T^{\rm dec}_{\nu_R}) = 1.66 \sqrt{g (T^{\rm dec}_{\nu_L}) } \
\frac{(T^{\rm dec}_{\nu_R})^2}{M_{\rm
    Pl}} \ \left( \frac{3}{\Delta
  N_\nu} \right)^{3/8} .
\label{Hubble}
\end{equation}
Substituting (\ref{Gamma}) and (\ref{Hubble}) into (\ref{haim1}) we obtain
\begin{equation}
\frac{\overline g}{M_{Z'}} = \left( \frac{3}{\Delta N_\nu}\right)^{3/32}
\left(\frac{13.28 \ \sqrt{g(T_{\nu_L}^{\rm dec})}}{M_{\rm Pl} \ {\cal
      K} \ (T_{\nu_R}^{\rm dec})^3} \right)^{1/4} \, .
\label{Tolo}
\end{equation}
For a given value of $\Delta N_\nu$, (\ref{Tolo}) conveniently yields a straight line 
plot of $\bar g~{\it vs.}~M_{Z'}$. In Fig.~\ref{fig:ggi_3} we provide graphs 
corresponding to the central value and $1\sigma$ limits for the values
of $\Delta N_\nu$ given in (\ref{deltanu}). The hatched region between
the highest and lowest lines reprersents the $\overline g-M_{Z'}$
parameter space consistent with the $\Delta N_\nu$ measurement.

To illustrate we calculate $\overline g$ for two candidate models. The
first is a set of variations on D-brane constructions which do not have
coupling constant unification. The second are two $U(1)$ models
($U(1)_\psi$ and $U(1)\chi$) which are embedded in a grand unified
exceptional $E_6$ group, with breaking pattern 
\begin{equation}
E_6 \to SO(10) \times
U(1)_\psi \to SU(5) \times U(1)_\psi \times U(1)_\chi \, .
\end{equation}
The latter two are interesting because they provided a test basis for
$Z'$ searches at ATLAS~\cite{:2012hf} and CMS~\cite{Chatrchyan:2012it}. For each of the $E_6$ models we may
write $g_i$ in (\ref{gbarra}) as $g_i = g_0 Q_i$, where in conformity
with grand unification we follow~\cite{Barger:2003zh} and choose
\begin{equation}
g_{0} = \sqrt{\frac{5}{3}} \ g_2  \ \tan \theta_W \sim 0.46 \,,
\end{equation}
with $g_2$ the $SU(2)_L$ coupling.  The charges $Q_i$ for the
different fermions are conveniently tabulated in~\cite{Barger:2003zh}.

In the D-brane construction, the Weyl fermions live at the brane
intersections of a particular 4-stack quiver configuration: $U(3)_C
\times SU(2)_L \times U(1)_{I_R} \times
U(1)_L$~\cite{Cremades:2003qj}.  The resulting $U(1)$ content gauges
the baryon number $B$ [with $U(1)_B \subset U(3)_C$], the lepton
number $L$, and a third additional abelian charge $I_R$ which acts as
the third isospin component of an $SU(2)_R$.  Contact with gauge
structures at TeV energies is achieved by a field rotation to couple
diagonally to hypercharge $Y$. Two of the Euler angles are determined
by this rotation and the third one is chosen so that one of the $U(1)$
gauge bosons couples only to an anomaly free linear combination of
$I_R$ and $B-L$~\cite{Anchordoqui:2012wt}.  Of the three original
abelian couplings, the baryon number coupling is fixed to be
$\sqrt{1/6}$ of the QCD coupling at the string scale. The orthogonal
nature of the rotation imposes one additional constraint on the
remaining couplings~\cite{Anchordoqui:2011eg}.  Since one of the two
extra gauge bosons is coupled to an anomalous current, its mass is
${\cal O}(M_s)$, as generated through some St\"uckelberg
mechanism. The other gauge boson is coupled to an anomaly free current
and therefore (under certain topological conditions) it can remain
massless and grow a TeV-scale mass through ordinary Higgs
mechanisms~\cite{Cvetic:2011iq}. We consider two extreme possibilities
in which the TeV-scale $Z'$ gauge boson is mostly $I_R$ or mostly
$B-L$. The chiral couplings ($g_i$) of these gauge boson are tabulated
in~\cite{Anchordoqui:2012wt}. We also consider a D-brane construct
with TeV-scale string compactification (the chiral couplings of this
model are given in Table IV of~\cite{Anchordoqui:2011eg}).  Details of
these assignments are given in the figure caption.  Termination of the
lines on the left reflects the LHC experimental limits on the mass of
the gauge boson, using null signals for enhancements in
dilepton~\cite{:2012hf,Chatrchyan:2012it} and
dijet~\cite{atlas2,:2012yf} searches.

The cosmology results from the Planck satellite would allow
determination of $N_{\rm eff}$ with a standard deviation of about
0.2~\cite{Hamann:2007sb,Galli:2010it}, whereas the future Large
Synoptic Survey Telescope (LSST) could determine $N_{\rm eff}$ with a
standard deviation of about 0.1~\cite{Joudaki:2011nw}. With this
enhanced sensitivity the hatched region will collapse to a line and
intersect for any given model its horizontal curve at the mass of the
$Z'$.

{\bf Note added}: After this work was finished a paper appeared on the
arXiv with a comprehensive study on dark radiation of $E_6$
models~\cite{SolagurenBeascoa:2012cz}. Our results are completely
consistent with those of Ref.~\cite{SolagurenBeascoa:2012cz}.

 We thank the Galileo Galilei Institute for Theoretical Physics for
 the hospitality and the INFN for partial support during the
 completion of this work. L.A.A.\ is supported by the U.S. National
 Science Foundation (NSF) under CAREER Grant PHY-1053663.  H.G.\ is
 supported by NSF Grant PHY-0757959.  G.S. is supported by the
 Department of Energy (DOE) Grant DE-FG02-91ER40690. Any opinions,
 findings, and conclusions or recommendations expressed in this
 material are those of the authors and do not necessarily reflect the
 views of NSF or DOE.

\end{document}